\documentclass[12pt]{article}
\usepackage{qJen}
\input{MyQcircuit.sty}
\begin{document}
\title{Quantum Circuit For\\
 Discovering from Data\\
 the Structure of\\
Classical Bayesian Networks}

\author{Robert R. Tucci\\
        P.O. Box 226\\
        Bedford,  MA   01730\\
        tucci@ar-tiste.com}

\date{\today}
\maketitle
\vskip2cm
\section*{Abstract}
We give some quantum circuits for calculating
the probability $P(G|D)$ of a
graph $G$ given data $D$. $G$ together
with a transition probability matrix
for each node of the graph, constitutes a
Classical Bayesian Network, or CB net for short.
Bayesian methods for calculating $P(G|D)$
have been given before (the so called
structural modular and ordered modular
models), but these earlier methods were designed to work on a classical computer. The goal of this paper is to
``quantum computerize" those earlier methods.

\newpage

\section{Introduction}

In this paper, we give some quantum circuits for calculating
the probability $P(G|D)$ of a
graph $G$ given data $D$. $G$ together
with a transition probability matrix
for each node of the graph, constitutes a
Classical Bayesian Network, or CB net for short.
Bayesian methods for calculating $P(G|D)$
have been given before (the so called
structural modular and ordered modular
models), but these earlier methods were designed to work on a classical computer. The goal of this paper is to
``quantum computerize" those earlier methods.

Often
in the literature, the word ``model"
  is used synonymously with ``CB net"
  and the word ``structure" is used synonymously
  with the bare ``graph" $G$,
  which is the CB net without the
  associated transition probabilities.

  The Bayesian methods
  for calculating $P(G|D)$
  that we will discuss in this paper
  assume a ``meta" CB net
  to predict $P(G|D)$ for a CB net with graph $G$. The meta CB nets usually assumed have a ``modular" pattern.
  Two
  types of modular meta CB nets
  have been studied in the literature.
  We will call them in this paper
  unordered modular and ordered modular
  models although unordered modular models are more commonly called
  structural modular models.

  Calculations with unordered modular models require that
 sums $\sum_G$
over graphs $G$ be performed.
Calculations with  ordered modular models require that,
besides sums $\sum_G$, sums $\sum_\sigma$
over ``orders" $\sigma$
be performed.
In some methods these two types of sums are
performed deterministically; in others, they
are both performed
by doing MCMC sampling of a probability distribution. Some hybrid methods
perform some of those sums deterministically and others
by sampling.

One of the first papers to
propose unordered modular models
appears to be Ref.\cite{Co-He}
by Cooper and Herskovits.
Their paper proposed
performing the $\sum_G$
by sampling.

One of the first papers to
propose
ordered modular models
appears to be
Ref.\cite{Fr-Ko}
by Friedman and Koller.
Their paper proposed
performing both  $\sum_G$
and  $\sum_\sigma$
by sampling.

Later on,
Refs.\cite{Ko-So,Ko}
by Koivisto and Sood
proposed a way
of doing $\sum_G$
deterministically using a technique
they call fast Mobius transform,
and performing
$\sum_\sigma$ also
deterministically by
using a technique they call
DP (dynamic programming).

Since the initial work of
Koivisto and Sood, several
workers (see, for example,
 Refs.\cite{Wo,He}) have proposed
hybrid methods that
use both sampling and
the deterministic methods
of Koivisto  and Sood.

So how can one quantum computerize
to some extent
the earlier classical computer methods
for calculating $P(G|D)$? One
partial way is to
replace sampling with classical
computers by sampling with
quantum computers (QCs).
An algorithm for sampling
CB nets
on a QC has been proposed by
Tucci in Ref.\cite{quibbs}.
A second possibility is to
replace the deterministic
summing of $\sum_G$ or $\sum_\sigma$  by quantum
summing of the style
discussed in Refs.\cite{qSym,qMob, afga},
wherein one uses a Grover-like algorithm
and the technique of
targeting two hypotheses.
This second possibility is what will be discussed in this paper,
for both types of modular models.

Finally, let us mention that some earlier
papers (see,
for example, Refs.\cite{Neven, Void, Microsoft}) have proposed using a quantum computer
to do AI related calculations
 reminiscent of the ones being tackled in this paper. However,
the methods proposed in those papers differ
greatly from the one in this paper.
Those papers
 either don't use Grover's algorithm, or if they do, they don't
use our techniques of targeting two hypotheses and blind targeting.

This paper assumes that
the reader has already read
most of Refs.\cite{qSym}
and \cite{qMob}.
by Tucci. Reading those
previous 2 papers is essential
to understanding this one
because
this paper applies
techniques described in
those 2 previous papers.

\label{sec-intro}
\section{Notation and Preliminaries}
Most of the notation that will be
used in this paper has already been
explained in previous papers by Tucci.
See, in particular, Sec.2
(entitled ``Notation and Preliminaries") of Refs.\cite{qSym} and \cite{qMob}.
In this section, we will discuss some
notation and definitions that
will be used in this paper but
which were not discussed in those
two earlier papers.

We will underline random variables.
For example,
we will say that the
random variable $\rvx$ has probability
distribution $P_\rvx(x)=P(\rvx=x)=P(x)$ and takes on values in the set $S_\rvx=val(\rvx)$. We
will sometimes also use $N_\rvx=|S_\rvx|$.

Throughout this paper, the symbol $n$,
if used as a scalar, will always denote the number of  nodes of a graph. However, $n$
will sometimes stand for the operator
$n=\ket{1}\bra{1}$ that measures the number of particles, either 0 or 1, in
a single qubit state. It will
usually be clear from context whether $n$
refers to the number of nodes or the number operator. In cases where we are using both meanings at the same time,
we will indicate the number operator by $n_{op}$ or $\rvn$ or $\hat{n}$
and the number of nodes simply by $n$.

As usual, an
ordered set or n-tuple (resp., unordered set)
will be indicated by
enclosing its elements with parentheses
(resp., braces)

We will use two dots between two
integers to denote all intervening integers. For example,
$(2..6)=(2,3,4,5,6)$,
$(6..2)=(6,5,4,3,2)$,
$\{2..6\}=\{6..2\}=\{2,3,4,5,6\}$.

We will use a backslash
to denote the exclusion of
the elements following the backslash in
an ordered or unordered set. For example,
$(2..6\bs 4,2)=(3,5,6)$,
$(6..2\bs 4,2)=(6,5,3)$,
and
$\{2..6\bs 4,2\}=
\{6..2\bs 4,2\}=\{3,5,6\}$.

We will use the
symbols $<,\leq ,>, \geq$
inside ordered or unordered sets
to denote various bounded sequences of
integers. For example,
$(<4)_1 =(1,2,3)$,
$(>4)_8 =(5,6,7,8)$,
$\{<4\}_1 =\{1,2,3\}$,
$\{>4\}_8 =\{5,6,7,8\}$,
$(\leq 4)_1 =(1,2,3,4)$, etc.

On occasion, we will use Stirling's approximation:

\beq
n! \approx
(2 \pi n)^{\frac{1}{2}}
\left(\frac{n}{e}\right)^n
\;.
\eeq
Note that $2^n< n!< n^n = 2^{n\log_2 n}$.

\section{Review of Classical Theory}
In this section, we will review
some previous theory
by other workers
(references already cited in  Sec.\ref{sec-intro}).
This previous theory
is the foundation of some
algorithms for
using {\it classical computers}
to discover the structure
of CB nets from data.
The theory defines
two types of ``modular models",
either
unordered or ordered.

\subsection{The Graph Set $\calb_n$ and its subsets}

The main goal of
the theory of modular models
is to give a ``meta" CB net that helps us to
discover the graph
of a specific CB net.
Before embarking on a
detailed  discussion of
 modular models, it is convenient to
discuss various sets of graphs.

The structure of an $n$-node\footnote{We will use the words ``vertex" and ``node" interchangeably.} ``bi-directed" graph $G$
is fully specified by giving,
 for each node $j\in\{0..n-1\}$
 of graph $G$, the set $pa_j=pa(j,G)\subset\{0..n-1\}$
of parents of node $j$.
Hence, we will make the following identification:
\beq
G=(pa_{n-1},\ldots,pa_1,pa_0)
\in\calb_n
\;,
\eeq
where

\beq
\calb_n = \underbrace{2^{\set{0..n-1}}\times\ldots
\times 2^{\set{0..n-1}}
\times 2^{\set{0..n-1}}}_{n\mbox{ times}}=(2^{\set{0..n-1}})^n
\;.
\eeq
Note that $|\calb_n|=2^{n^2}$.

Suppose $G,G'\in \calb_n$.
Then we will write
$G\subset G'$ if $pa(j, G)\subset pa(j,G')$
for all $j$.

As is customary in the literature,
we will abbreviate the phrase ``Directed
Acyclic Graph" by DAG.
Define

\beqa
\DAG_n &=&\{G\in\calb_n : G\mbox{ is \DAG}\}
\\
&=&
2^{\set{n-2..0}}
\times\ldots
\times 2^{\set{1,0}}
\times 2^{\set{0}}
\times 2^{\emptyset}
\;.
\eeqa
Note that
$
|\DAG_n| =
2^{n-1}
\ldots
2^{2}
 2^{1}
2^{0}=
2^{\frac{n(n-1)}{2}}
$.
A special element of $\DAG_n$ is the
Fully Connected Graph with $n$ nodes,
defined by

\beqa
FCG_n &=& (\set{<n-1},\ldots,\set{<2},\set{<1},\set{<0})\\
&=&(\set{n-2..0},\ldots,\set{1,0},\set{0},\emptyset)
\;.
\eeqa
Note that $\DAG_n=\set{G\in \calb_n: G\subset FCG_n}$.

As an example,
consider $\DAG_n$ for
$n=3$. One has
$|\DAG_3|=8$. To list
the elements of $\DAG_3$,
one begins by noticing
that any $G=(pa_2,pa_1,pa_0)\in \DAG_3$
must have

\beq
\begin{array}{r}
pa_0\in \set{\emptyset}
\\
pa_1\in \set{\set{0},\emptyset}
\\
pa_2\in \set{\set{1,0},\set{1},\set{0},\emptyset}
\end{array}
\;.
\eeq
Thus, we get
the following
list of elements of $\DAG_3$:

\beq
\begin{array}{|ll|}\hline
G_0=(\emptyset,\emptyset,\emptyset)
&
\begin{array}{c}
\xymatrix@R=1pc{
&1&
\\
2 & &0
}
\end{array}
\\  \hline
G_1=(\emptyset,\set{0},\emptyset)
&
\begin{array}{c}
\xymatrix@R=1pc{
&1&
\\
2 & &0\ar[ul]
}
\end{array}
\\  \hline
G_2=(\set{1},\emptyset,\emptyset)
&
\begin{array}{c}
\xymatrix@R=1pc{
&1\ar[ld]&
\\
2 & &0
}
\end{array}
\\  \hline
G_3=(\set{1},\set{0},\emptyset)
&
\begin{array}{c}
\xymatrix@R=1pc{
&1\ar[ld]&
\\
2 & &0\ar[ul]
}
\end{array}
\\  \hline
G_4=(\set{0},\emptyset,\emptyset)
&
\begin{array}{c}
\xymatrix@R=1pc{
&1&
\\
2 & &0\ar[ll]
}
\end{array}
\\  \hline
G_5=(\set{0},\set{0},\emptyset)
&
\begin{array}{c}
\xymatrix@R=1pc{
&1&
\\
2 & &0\ar[ul]\ar[ll]
}
\end{array}
\\  \hline
G_6=(\set{1,0},\emptyset,\emptyset)
&
\begin{array}{c}
\xymatrix@R=1pc{
&1\ar[ld]&
\\
2 & &0\ar[ll]
}
\end{array}
\\  \hline
G_7=(\set{1,0},\set{0},\emptyset)
&
\begin{array}{c}
\xymatrix@R=1pc{
&1\ar[ld]&
\\
2 & &0\ar[ul]\ar[ll]
}
\end{array}
\\  \hline
\end{array}
\;
\eeq
Note that $FCG_3=G_7$.

\begin{figure}[h]
    \begin{center}
    \epsfig{file=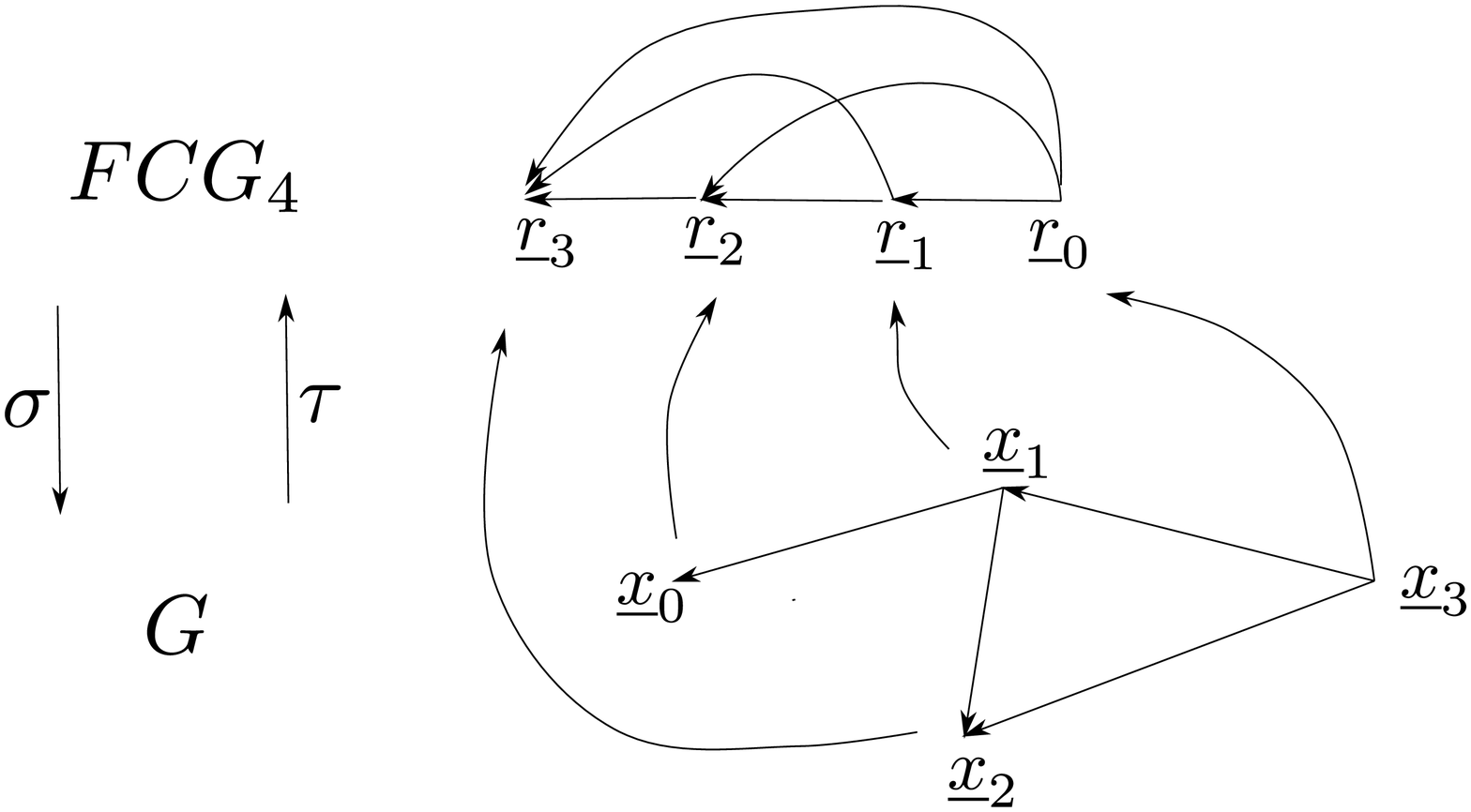, height=1.75in}
    \caption{The permutation
    $\sigma$ maps
    the nodes of the
    fully connected graph $FCG_4$
    to the nodes of the graph $G$.}
    \label{fig-fcg-g-map}
    \end{center}
\end{figure}

Most of the previous literature
speaks about an order or, more precisely,
a linear order among the graphs of
$\DAG_n$. In this paper,
instead of using the language of linear orderings, we chose
to speak in the totally equivalent language
of permutations $\sigma\in Sym_n$.

Let
$FCG_n\in \DAG_n$
be the unique fully connected
graph with $n$ nodes,
where $\rvr_0$ is the root node,
$\rvr_1$ is the
child of $\rvr_0$, $\rvr_2$ is
the child of $\rvr_1$ and $\rvr_0$, and so on.
Given any
$G\in \DAG_n$,
its nodes can be ordered
topologically.
This is
tantamount to finding a permutation $\sigma\in Sym_n$
mapping the nodes of $FCG_n$ to the nodes of $G$. Let $\tau=\sigma^{-1}$.
See Fig.\ref{fig-fcg-g-map}
for an example with
$n=4$. In that figure

\beq
\tau=
\left(
\begin{array}{cccc}
\rvx_0 & \rvx_1 & \rvx_2 & \rvx_3
\\
\rvr_2 & \rvr_1 & \rvr_3 & \rvr_0
\end{array}
\right)
\;
\eeq
so $\rvx_0^\tau=
\rvx_{0^{\tau}}=\rvr_2$, etc.
Note that in that figure,
the parent sets of each node
of $G$
are related to those of $FCG_4$ as follows:

\beq
\begin{array}{l|l}
pa(3,G)=\emptyset
&
pa(3^\tau, FCG_4)=\emptyset
\stackrel{\sigma}{\longrightarrow}
\emptyset
\\
pa(2,G)=\{1,3\}
&
pa(2^\tau, FCG_4)=\{0,1,2\}
\stackrel{\sigma}{\longrightarrow}
\{3,1,0\}
\\
pa(1,G)=\{3\}
&
pa(1^\tau, FCG_4)=\{0\}
\stackrel{\sigma}{\longrightarrow}
\{3\}
\\
pa(0,G)=\{1\}
&
pa(0^\tau, FCG_4)=\{0,1\}
\stackrel{\sigma}{\longrightarrow}
\{3,1\}
\end{array}
\;.
\label{eq-pa-sets-eg}
\eeq
Eq.(\ref{eq-pa-sets-eg})
implies
that

\beq
pa_j = pa(j, G)
\subset [pa(j^\tau, FCG_n)]^\sigma
=
\set{<j^\tau}^\sigma
\;.
\eeq

For each $\sigma\in Sym_n$, define
the graph  $FCG_n^\sigma$ by

\beq
FCG_n^\sigma = (pa_j)_{\forall j}
\mbox{ , where } pa_j = \set{<j^\tau}^\sigma
\;
\eeq
Henceforth, we will say that
a graph
$G\in \DAG_n$ is consistent with permutation $\sigma\in Sym_n$
if $G$ can be obtained by
erasing some arrows from $FCG_n^\sigma$.
Equivalently
$G$ consistent with $\sigma$ if
$G\subset FCG_n^\sigma$.
Define

\beqa
(\DAG_n)_\sigma&=&
\{
G\in \DAG_n: G\subset FCG_n^\sigma\}
\\
&=&
2^{\set{<(n-1)^\tau}^\sigma}
\times \ldots \times
2^{\set{<2^\tau}^\sigma}
\times
2^{\set{<1^\tau}^\sigma}
\times
2^{\set{<0^\tau}^\sigma}
\;.
\eeqa
Note $|(\DAG_n)_\sigma|=|\DAG_n|$.
Another useful
set to consider is

\beq
(Sym_n)_G=
\{
\sigma\in Sym_n: G\subset FCG_n^\sigma\}
\;.
\eeq
Note that whereas $|Sym_n|=n!$, $|(Sym_n)_G|$ is much harder to calculate for large $n$.

As an example,
let us calculate
$FCG_n^\sigma$
for $n=3$
and all $\sigma\in Sym_3$.
One finds

\beq
\begin{array}{|c|c|c|c|c|c|}
\hline
\sigma
&
\tau=\sigma^{-1}
&\s
(\set{<2}^\sigma,
\set{<1}^\sigma,
\set{<0}^\sigma)
&
\s
(\set{<2^\tau}^\sigma,
\set{<1^\tau}^\sigma,
\set{<0^\tau}^\sigma)
&
FCG_3^\sigma
\\\hline\hline
\sigma_0=
\left(
\begin{array}{ccc}
0&1&2
\\
0&1&2
\end{array}
\right)
&
\left(
\begin{array}{ccc}
0&1&2
\\
0&1&2
\end{array}
\right)
&
(\set{1,0},\set{0},\emptyset)
&
\begin{array}{l}
\s(\set{<2}^\sigma,\set{<1}^\sigma,\set{<0}^\sigma)
\\
=(\set{1,0},\set{0},\emptyset)
\end{array}
&
\begin{array}{c}
\xymatrix@R=.5pc@C=.5pc{
&\s 1\ar[ld]&
\\
\s 2 & &\s 0\ar[ul]\ar[ll]
}
\end{array}
\\ \hline
\sigma_1=
\left(
\begin{array}{ccc}
0&1&2
\\
0&2&1
\end{array}
\right)
&
\left(
\begin{array}{ccc}
0&1&2
\\
0&2&1
\end{array}
\right)
&
(\set{2,0},\set{0},\emptyset)
&
\begin{array}{l}
\s(\set{<1}^\sigma,\set{<2}^\sigma,\set{<0}^\sigma)
\\
=(\set{0},\set{2,0},\emptyset)
\end{array}
&
\begin{array}{c}
\xymatrix@R=.5pc@C=.5pc{
&\s 1&
\\
\s 2\ar[ur] & &\s 0\ar[ul]\ar[ll]
}
\end{array}
\\ \hline
\sigma_2=
\left(
\begin{array}{ccc}
0&1&2
\\
1&0&2
\end{array}
\right)
&
\left(
\begin{array}{ccc}
0&1&2
\\
1&0&2
\end{array}
\right)
&
(\set{0,1},\set{1},\emptyset)
&
\begin{array}{l}
\s(\set{<2}^\sigma,\set{<0}^\sigma,\set{<1}^\sigma)
\\
=(\set{1,0},\emptyset,\set{1})
\end{array}
&
\begin{array}{c}
\xymatrix@R=.5pc@C=.5pc{
&\s 1\ar[ld]\ar[rd]&
\\
\s 2 & &\s 0\ar[ll]
}
\end{array}
\\ \hline
\sigma_3=
\left(
\begin{array}{ccc}
0&1&2
\\
2&0&1
\end{array}
\right)
&
\left(
\begin{array}{ccc}
0&1&2
\\
1&2&0
\end{array}
\right)
&
(\set{0,2},\set{2},\emptyset)
&
\begin{array}{l}
\s(\set{<0}^\sigma,\set{<2}^\sigma,\set{<1}^\sigma)
\\
=(\emptyset,\set{2,0},\set{2})
\end{array}
&
\begin{array}{c}
\xymatrix@R=.5pc@C=.5pc{
&\s 1&
\\
\s 2\ar[rr]\ar[ru] & &\s 0\ar[ul]
}
\end{array}
\\ \hline
\sigma_4=
\left(
\begin{array}{ccc}
0&1&2
\\
1&2&0
\end{array}
\right)
&
\left(
\begin{array}{ccc}
0&1&2
\\
2&0&1
\end{array}
\right)
&
(\set{2,1},\set{1},\emptyset)
&
\begin{array}{l}
\s(\set{<1}^\sigma,\set{<0}^\sigma,\set{<2}^\sigma)
\\
=(\set{1},\emptyset,\set{2,1})
\end{array}
&
\begin{array}{c}
\xymatrix@R=.5pc@C=.5pc{
&\s 1\ar[ld]\ar[rd]&
\\
\s 2\ar[rr] & &\s 0
}
\end{array}
\\ \hline
\sigma_5=
\left(
\begin{array}{ccc}
0&1&2
\\
2&1&0
\end{array}
\right)
&
\left(
\begin{array}{ccc}
0&1&2
\\
2&1&0
\end{array}
\right)
&
(\set{1,2},\set{2},\emptyset)
&
\begin{array}{l}
\s(\set{<0}^\sigma,\set{<1}^\sigma,\set{<2}^\sigma)
\\
=(\emptyset,\set{2},\set{2,1})
\end{array}
&
\begin{array}{c}
\xymatrix@R=.5pc@C=.5pc{
&\s 1\ar[rd]&
\\
\s 2 \ar[rr]\ar[ur]
& &\s 0
}
\end{array}
\\ \hline
\end{array}
\;.
\label{eq-fcg-3-sig}
\eeq

Note that for each $\sigma\in Sym_3$,
$\DAG_3^\sigma$ are all
graphs $G$ such that $G\subset FCG_3^\sigma$
where $FCG_3^\sigma$ is given by last column of Eq.(\ref{eq-fcg-3-sig}).

Henceforth, whenever we
write
a sum over $G$ without specifying the range of the sum, we will mean
over the range $\calb_n$.
Likewise, a sum over permutations $\sigma$ without specifying the range
should be interpreted as being over the range $Sym_n$. Thus,

\beq
\sum_G\sum_\sigma=
\sum_{G\in \calb_n}\sum_{\sigma\in Sym_n}
\;.
\eeq

Any set $\calf\subset\calb_n$ will be called a {\bf feature set}.
Given any probability distribution
$P(G)$ for $G\in \calb_n$, define
$P(\calf)$ by

\beq
P(\calf)=\sum_{G\in \calf}P(G)=
\sum_{G\in \calb_n} I_\calf(G) P(G)
\;,
\eeq
where $I_\calf(G)=\theta(G\in\calf)$
is an indicator function.

A set $\calf\subset \calb_n$
is said to be a {\bf modular
feature set} if it equals a cartesian product

\beq
\calf=\calf_{n-1}\times \ldots
\calf_1\times\calf_0 \subset \calb_n
\;.
\eeq
We will sometimes
denote such a set by $\calf = \bigotimes_j \calf_j$. Note that for a modular feature set,

\beq
1_\calf(G)=
\prod_j 1_{\calf_j}(pa_j)
\;.
\eeq
As an example, for any two nodes $j_1,j_2$,
the feature set for the
edge $j_1\rarrow j_2$ is
$\calf = \bigotimes_j \calf_j$,
where

\beq
\begin{array}{l}
\calf_{j_2}=\set{pa_{j_2}: pa_{j_2}\subset \set{0..n-1},j_1\in pa_{j_2}}
\\
\mbox{ for }j\neq j_2, \;\;\;\calf_j=\set{pa_{j}:pa_{j}\subset \set{0..n-1}}
\end{array}
\;.
\eeq

\subsection{Unordered Modular Models}
\label{sec-class-unord}

\begin{figure}[h]
    \begin{center}
    \epsfig{file=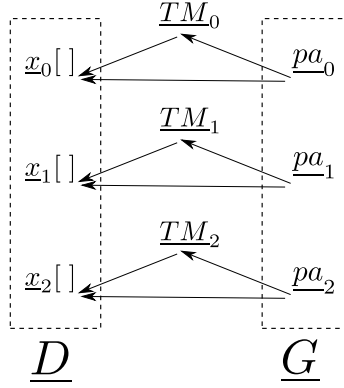, height=2in}
    \caption{Meta CB net for unordered modular model with
    $n=3$.}
    \label{fig-unord-modular}
    \end{center}
\end{figure}

In this section, we will discuss
the meta CB net which defines
unordered modular models.

The meta CB net for unordered modular models
is illustrated by Fig.\ref{fig-unord-modular}
 for the case of 3 nodes, $n=3$.

We will assume that the random
variable $\ul{G}$ takes on values:

 \beq
 G=(pa_j)_{\forall j}\in \calb_n
 \;.
 \eeq

Let index $m\in \{0..M-1\}$
label measurements and index $j\in\{0..n-1\}$
label nodes. Let $x_j[m]\in S_{\rvx_j}$ for all $j$ and $m$. Let
$x_j\sq  = \{x_j[m]\}_{\forall m}$,
$x^n\sq  =\{x_j\sq \}_{\forall j}=\{x_j[m]\}_{\forall j, m}$.
We will assume that the random variable $\ul{D}$ takes on values

\beq
D = x^n\sq\in S_{\rvx_0}^M\times
S_{\rvx_1}^M\times\ldots\times S_{\rvx_{n-1}}^M
\;.
\eeq
$\ul{D}$ is the data from which
we intend to infer the structure $\ul{G}$ of a
CB net.

We will assume that
$P(pa_j)$ is proportional to
$\theta(pa_j\subset \set{<j})$
(in other words, that its
support is inside $2^{\set{<j}}$)
so that

\beq
\sum_{pa_j\subset \set{<j}}P(pa_j)=1
\;.
\eeq

 Let
\beq
P(G) = \prod_j \left\{
\theta(pa_j\subset \set{<j})P(pa_j)
\right\}
\;.
\label{eq-pg-prod}
\eeq
Note that this implies that $P(G)$ is proportional to
$\theta(G\subset FCG_n)$.
In light of our assumption
about the support of $P(pa_j)$,
there is no need to write down the
$\theta(pa_j\subset \set{<j})$
in Eq.(\ref{eq-pg-prod}),
but we will write it as a reminder.

Let
\beq
P(D|G)=
\prod_j
P(x_j\sq |pa_j)
\;
\eeq
where

\beq
P(x_j\sq |pa_j)=
\sum_{TM_j}
P(x_j\sq |pa_j, TM_j)P(TM_j|pa_j)
\;.
\label{eq-xj-paj}
\eeq
Let $TM^n = (TM_0, TM_1, \ldots, TM_{n-1})$.
The role
 of the variable $TM_j$
 is to parameterize the transition matrix for node $j$.
Thus, summing over $TM_j$ is
equivalent to summing over all possible
transition matrices for node $j$.

The
$P(x_j\sq |pa_j)$
can be modelled
by a reasonable probability distribution.
For example,
under some reasonable assumptions,
Cooper and Herskovits
find in Ref.\cite{Co-He} that

\beq
P(x_j\sq |pa_j)=
\frac{[N_{\rvx_j}-1]!}
{[\sum_{x_j}\left\{N(j,x_j,pa_j)\right\}+ N_{\rvx_j}-1]!}
\prod_{x_j\in S_{\rvx_j}}
\left\{N(j,x_j, pa_j)!\right\}
\;,
\eeq
where

\beq
N(j, x_j, pa_j)=
\sum_{m=0}^{M-1}
\theta(x_j[m]=x_j)\theta(pa_j[m]=pa_j)
\;,
\eeq
and $N_{\rvx_j} = |S_{\rvx_j}|$.
This is a special case of
the Dirichlet probability
distribution.

Now that our meta CB net
for unordered modular models is fully
defined, we can calculate the $P(G|D)$
it predicts.

Let
\beq
\beta_j(pa_j)=P(x_j\sq|pa_j)P(pa_j)
\;.
\eeq
Then

\beqa
P(G|D) &=& \frac{P(D|G)P(G)}{\sum_G num}
\\
&=&
\frac{\prod_j \left\{\theta(pa_j\subset \set{<j}) \beta_j(pa_j)\right\}}
{\sum_G num}
\;,
\eeqa
where $\sum_G num$ means the numerator summed over $G$.
If $\calf=\bigotimes_j \calf_j$ is a modular feature set,

\beq
P(\calf|D) =
\frac{\prod_j \left\{\sum_{pa_j}1_{\calf_j}(pa_j)
\theta(pa_j\subset \set{<j})\beta_j(pa_j)\right\}}
{(num)_{\calf\rarrow\calb_n}}
\;,
\label{eq-p-f-d-unord}
\eeq
where
$(num)_{\calf\rarrow\calb_n}$
means the numerator with $\calf$
replaced by $\calb_n$.

\subsection{Ordered Modular Models}
\label{sec-class-ord}

\begin{figure}[h]
    \begin{center}
    \epsfig{file=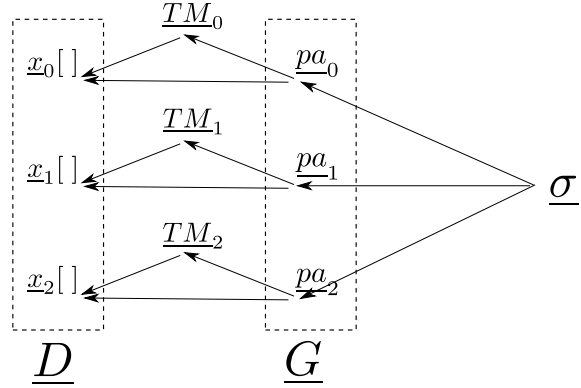, height=2in}
    \caption{Meta CB net for ordered modular model with
    $n=3$.}
    \label{fig-ord-modular}
    \end{center}
\end{figure}

In this section, we will discuss
the meta CB net which defines
ordered modular models.

The meta CB net for ordered modular models
is illustrated by Fig.\ref{fig-ord-modular}
 for the case of 3 nodes, $n=3$.
Comparing Figs.\ref{fig-unord-modular}
and \ref{fig-ord-modular},
we see that unordered modular models
presume that the
different nodes of the graph $G$
 we are trying to discover, are uncorrelated,
an unwarranted assumption
in some cases (for example, if two
nodes have a common parent). Ordered
modular models permit us to model some of the correlation between the nodes.

As for the unordered case
described in Sec.\ref{sec-class-unord},
we will assume that the random
variable $\ul{G}$ takes on values:

 \beq
 G=(pa_j)_{\forall j}\in \calb_n
 \;,
 \eeq
and the random variable $\ul{D}$ takes on values

\beq
D = x^n\sq\in S_{\rvx_0}^M\times
S_{\rvx_1}^M\times\ldots\times S_{\rvx_{n-1}}^M
\;.
\eeq

We will put
a line over the $P$ as in $\pp()$ for probability
distributions referring to the
ordered modular case to
distinguish them from
those referring to the unordered modular
case, which we will continue to
represent by
$P()$ without the overline.

We
will assume that $\pp(pa_j|\sigma)$ is proportional to
$\theta(pa_j\subset \set{<j^\tau}^\sigma)$
so that

\beq
\sum_{pa_j \subset \set{<j^\tau}^\sigma} \pp(pa_j|\sigma)=1
\;.
\eeq

Let

\beq
\pp(G|\sigma)=
\prod_j \pp(pa_j|\sigma)=
\prod_j\left\{
\theta(pa_j\subset \set{<j^\tau}^\sigma)\pp(pa_j|\sigma)
\right\}
\;.
\eeq
Note that this implies that
$\pp(G|\sigma)$ is proportional to
$\theta(G\subset FCG_n^\sigma)$.

Let
\beq
\pp(D|G)=
\prod_j
\pp(x_j\sq|pa_j)
\;,
\eeq
where

\beq
\pp(x_j\sq|pa_j)
=\sum_{TM_j}
\pp(x_j\sq|pa_j, TM_j)
\pp(TM_j|pa_j)
\;.
\eeq
As
in the unordered case described in Sec.\ref{sec-class-unord},
$\pp(x_j\sq|pa_j)$
can be modelled by
using a Dirichlet or some other reasonable probability distribution.

Now that our meta CB net
for ordered modular models is fully
defined, we can calculate the $\pp(G|D)$
it predicts.

Define
\beq
\bbeta_j(pa_j|\sigma)=\pp(x_j\sq|pa_j)\pp(pa_j|\sigma)
\;.
\eeq
Then

\beqa
\pp(G|D) &=& \frac{\sum_\sigma\pp(D|G)\pp(G|\sigma)\pp(\sigma)}{\sum_G num}
\\
&=&
\frac{\sum_\sigma \pp(\sigma)\prod_j \left\{\theta(pa_j\subset \set{<j^\tau}^\sigma) \bbeta_j(pa_j|\sigma)\right\}}
{\sum_G num}
\;.
\eeqa

Note that if

\begin{subequations}
\beq
\pp(x_j\sq|pa_j)= P(x_j\sq|pa_j)
\;,
\eeq

\beq
\pp(pa_j|id)=P(pa_j)\theta(pa_j\subset \set{<j})
\;,
\eeq
and

\beq
\pp(\sigma)=\delta(\sigma,id)
\;,
\eeq
\end{subequations}
where $id$ is the identity permutation, then

\beq
\pp(G|D) = P(G|D)
\;.
\label{eq-pp-to-p-1}
\eeq

If $\calf=\otimes_j \calf_j$ is a modular feature set,

\beq
\pp(\calf|D) =
\frac{\sum_\sigma\pp(\sigma)\prod_j \left\{\sum_{pa_j}1_{\calf_j}(pa_j)
\theta(pa_j\subset \set{<j^\tau}^\sigma)\bbeta_j(pa_j|\sigma)\right\}}
{(num)_{\calf\rarrow\calb_n}}
\;.
\eeq

Next we will assume that

\beq
\pp(pa_j|\sigma) = \pp(pa_j|\set{<j^\tau}^\sigma)
\;,
\;\;
\bbeta_j(pa_j|\sigma)=
\bbeta_j(pa_j|\set{<j^\tau}^\sigma)
\;
\eeq
When this is true, it is convenient to define the
following $h()$ functions for all
$j\in\set{0..n-1}$ and
$\sigma\in Sym_n$:

\beq
h({j^\sigma}|\set{<j}^\sigma)=
\sum_{pa_{j^\sigma}}1_{\calf_{j^\sigma}}(pa_{j^\sigma})
\theta(pa_{j^\sigma}\subset \set{<j}^\sigma)\bbeta_{j^\sigma}(pa_{j^\sigma}|\set{<j}^\sigma)
\;.
\eeq
$\pp(\calf|D)$ can be expressed in
terms of the
$h()$ functions as follows:

\beq
\pp(\calf|D) =
\frac{\sum_\sigma\pp(\sigma)\prod_j h(j^\sigma|\set{<j}^\sigma)}
{(num)_{\calf\rarrow\calb_n}}
\;.
\label{eq-p-f-d-ord-long}
\eeq

Suppose that for any $\sigma\in\pi_{Sym_n}$,

\beq
\pp(\sigma) =
\frac{
\prod_{j=0}^{n-1} \Phi(j^\sigma|\set{<j}^\sigma)}
{\sum_\sigma num}
\;
\label{eq-special-p-sigma}
\eeq
where $\Phi(j|S)$
is some non-negative function
defined
for all $j\in\set{0..n-1}$
and $S\subset\set{0..n-1\backslash j}$.

A completely general $\pp(\sigma)$
has $n!-1=\calo(n^n)$ real degrees of freedom.
On the other hand, the special $\pp(\sigma)$
given by Eq.(\ref{eq-special-p-sigma})
has $n2^{n-1}-1$
degrees of freedom so it is just a
poor facsimile of
the full  $\pp(\sigma)$.
Nevertheless, this
special $\pp(\sigma)$
is a nice bridge function between
two interesting extremes:
If  $\Phi()$ is
a constant function,
then $\pp(\sigma)$ is too.
Furthermore, if
$\Phi(j|S)= \delta(\set{<j},S)$,
then it is easy to see that
$\pp(\sigma) = \delta(\sigma, id)$,
where $id$ is the identity permutation.
But $\pp(\sigma) = \delta(\sigma, id)$ is just
the unordered modular model.
We conclude that
Eq.(\ref{eq-special-p-sigma})
includes the uniform distribution
and the unordered modular model as
special cases.

Another nice feature of the special
$\pp(\sigma)$ of Eq.(\ref{eq-special-p-sigma}) is that
for each $j$, we can
redefine
$h(j^\sigma|\set{<j}^\sigma)$
so that it absorbs the corresponding
$\Phi(j^\sigma|\set{<j}^\sigma)$ function.
Thus, if we assume the special $\pp(\sigma)$,
then, without further loss of generality,
Eq.(\ref{eq-p-f-d-ord-long}) becomes

\beq
\pp(\calf|D) =
\frac{\sum_\sigma\prod_j h(j^\sigma|\set{<j}^\sigma)}
{(num)_{\calf\rarrow\calb_n}}
\;.
\label{eq-p-f-d-ord}
\eeq

For instance, when
$n=3$,
we have

\beq
\pp(\calf| D)=
\frac{\sum_\sigma
h(2^\sigma| \{ 1,0\}^\sigma)
h(1^\sigma|0^\sigma) h(0^\sigma)}
{(num)_{\calf\rarrow \calb_n}}
\;.
\eeq
Hence

\beq
\left.
\begin{array}{c}
\left.
\begin{array}{c}
h_{2|\{1,0\}}h_{1|0}h_0
\\
h_{2|\{0,1\}}h_{0|1}h_1
\end{array}
\right\}
A = h_{2|\{1,0\}}
(h_{1|0}h_0 + h_{0|1}h_1)
\\
\left.
\begin{array}{c}
h_{1|\{0,2\}}h_{0|2}h_2
\\
h_{1|\{2,0\}}h_{2|0}h_0
\end{array}
\right\}
B = h_{1|\{0,2\}}
(h_{0|2}h_2 + h_{2|0}h_0)
\\
\left.
\begin{array}{c}
h_{0|\{2,1\}}h_{2|1}h_1
\\
h_{0|\{1,2\}}h_{1|2}h_2
\end{array}
\right\}
C = h_{0|\{2,1\}}
(h_{2|1}h_1 + h_{1|2}h_2)
\end{array}
\right\}
\pp(\calf|D)=
\frac{A+B+C}{(num)_{\calf\rarrow \calb_n}}
\;.
\eeq

\section{Quantum Circuits for Calculating $P(\calf|D)$}
In this section,
 we will give
 quantum circuits for calculating $P(\calf|D)$
for
both unordered and ordered
modular models.
Two types of sums, sums $\sum_G$
over graphs,
and sums $\sum_\sigma$
over permutations, need to
be performed to calculate
$P(\calf|D)$.
The methods proposed in this section
perform both of these sums
using a Grover-like algorithm
in conjunction with the techniques of blind targeting and
targeting two hypotheses, in the style
discussed in Refs.\cite{qSym,qMob}
\begin{figure}[h]
 \begin{center}
    \epsfig{file=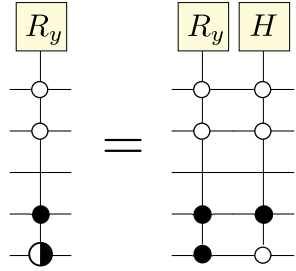, width=1.5in}
    \caption{Definition of
    halfmoon vertices used in Fig.\ref{fig-qJen-ckt}.
    }
    \label{fig-halfmoon}
    \end{center}
\end{figure}
\subsection{Unordered Modular Models}
In this section, we present
one possible method for calculating $P(\calf|D)$ for unordered modular models, where $P(\calf|D)$ is given by Eq.(\ref{eq-p-f-d-unord}).

One possible method of doing this is
as follows: for each $j\in\{0..n-1\}$,
calculate $\sum_{pa_j\subset\set{<j}}$
using the technique of Ref.\cite{qMob}
for calculating Mobius transforms.

\subsection{Ordered Modular Models}

\begin{figure}[h]
    \begin{center}
    \epsfig{file=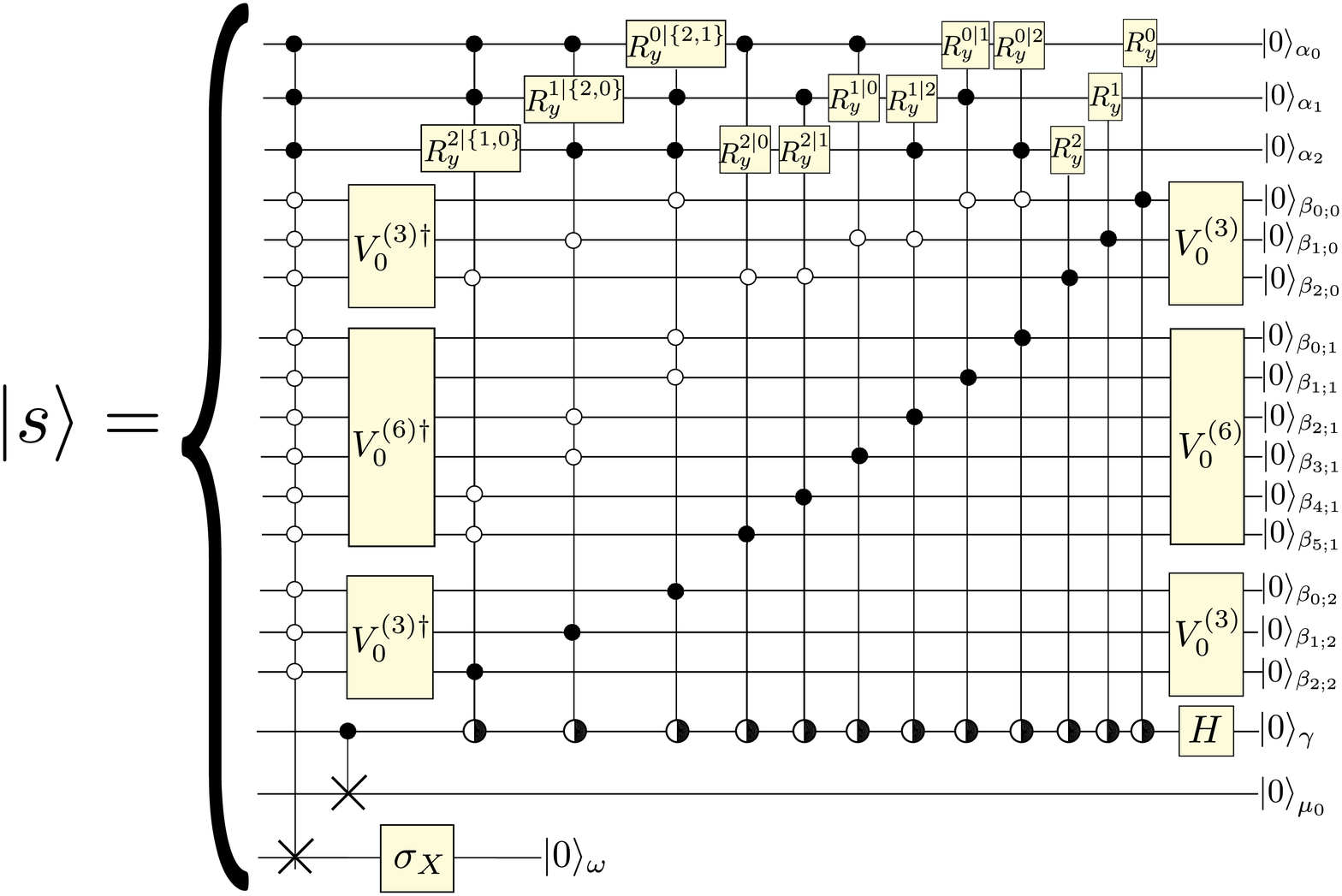, width=6in}
    \caption{Circuit for
    generating $\ket{s}$
    used in AFGA to calculate
    $\pp(\calf| D)$
    for ordered modular model.
    The $V_0^{(\lam)}$ matrices
    are defined in Ref.\cite{qSym}.
    }
    \label{fig-qJen-ckt}
    \end{center}
\end{figure}
In this section, we present
one possible method for calculating $\pp(\calf|D)$ for ordered modular models,
where $\pp(\calf|D)$ is given by Eq.(\ref{eq-p-f-d-ord}).
Eq.(\ref{eq-p-f-d-ord}) is just a sum over
all permutations in $Sym_n$.
We have already shown how to
do sums over permutations in Ref.\cite{qSym}.
So we could consider our job already done.
However, using the method of Ref.\cite{qSym}
would entail
the non-trivial task of
finding a way of compiling the argument
under the sum over permutations,
a complicated $n$-fold product of $h()$ functions.
In this section, we
will give a
method for calculating Eq.(\ref{eq-p-f-d-ord})
that does not require
compiling this
$n$-fold product of $h()$ functions.

For each $j\in \set{0..n-1}$
and $S\subset \{0..n-1\bs j\}$,
define

\beq
h(j|S) =\sin(\theta_{j|S})
\;
\eeq
and

\beq
R_y^{j|S}=
\exp( -i\sigma_Y \theta_{j|S})
\;.
\eeq
Let $|S|=\ell$.
Define

\beq
N_2(\beta_{;\ell})= n
\left(
\begin{array}{c}
n-1
\\
\ell
\end{array}
\right)
\;
\eeq
and

\beq
N_2(\beta)=
 \sum_{\ell=0}^{n-1}
N_2(\beta_{;\ell})=n2^{n-1}
\;.
\label{eq-exp-n2b}
\eeq
The following fraction
will also come into play:

\beq
\epsilon=
\frac{1}{
\prod_{\ell=0}^{n-1}
N_2(\beta_{;\ell})
}
\;.
\eeq

We will use qubits $\beta_{c;\ell}$
for $c\in\set{0..N_2(\beta_{;\ell})-1}$
and
$\ell\in\set{0..n-1}$.
Let $\beta_{;\ell}=(\beta_{c;\ell})_{\forall c}$
and
$\beta = (\beta_{;\ell})_{\forall \ell}$.

For concreteness,
we will use $n=3$ henceforth in this section,
but it will be obvious
how to draw
an analogous
circuit
for arbitrary $n$.

We want
all
horizontal lines
in Fig.\ref{fig-qJen-ckt}
to represent qubits.
Let $\alpha= \alpha^3$
and
$\beta=
(\beta_{;0},\beta_{;1},\beta_{;2})
=(\beta^3_{;0},\beta^6_{;1},\beta^3_{;2})
=\beta^{12}$.

Define

\beq
T_x(\alpha)=
\epsilon
\sum_{j_0=0}^{2}
\sum_{j_1=0}^{2}\theta(j_1\neq j_0)
\sum_{j_2=0}^{2}\theta(j_2\neq j_1,j_2 \neq j_0)
T_x[j_2,2](\alpha)
T_x[j_1,1](\alpha)
T_x[j_0,0](\alpha)
\;
\eeq
for $x=0,1$, where

\beq
T_1[j,\ell](\alpha)=
\prod_{S\in\set{S: S\subset \set{0..n-1\bs j}, |S|=\ell}}
[e^{-i\sigma_Y(\alpha_j)\theta_{j|S}}]^
{\prod_{k\in S}P_1(\alpha_k)}
\;
\eeq
and

\beq
T_0[j,\ell](\alpha)=
\prod_{S\in\set{S: S\subset \set{0..n-1\bs j}, |S|=\ell}}
[H(\alpha_j)]^
{\prod_{k\in S}P_1(\alpha_k)}
\;
\eeq
where $H(\alpha_j)$ is a Hadamard matrix acting on bit $\alpha_j$.
Also define

\beq
\pi(\alpha)=
\prod_{j=0}^{2} P_1(\alpha_j)
\;,
\eeq
and

\beq
\pi(\beta)=
\left\{
\begin{array}{l}
\prod_{c=0}^2 P_0(\beta_{c;0})
\\
\prod_{c=0}^5 P_0(\beta_{c;1})
\\
\prod_{c=0}^2 P_0(\beta_{c;2})
\end{array}
\right.
\;.
\eeq

Our method
for
calculating
$\pp(\calf|D)$
consists of applying the algorithm
AFGA\footnote{As discussed
 in Ref.\cite{qSym},
 we recommend the AFGA
 algorithm, but Grover's original
 algorithm or any other
 Grover-like algorithm
 will also work
 here, as long as it
 drives a
 starting state $\ket{s}$
 to a target state $\ket{t}$.} of Ref.\cite{afga}
in the way that was described in
Ref.\cite{qSym},
using the techniques
of targeting two hypotheses
and blind targeting.
As in Ref.\cite{qSym},
when we apply AFGA in this section,
we will use a sufficient target $\ket{0}_\omega$.
All that remains for
us to do to
fully specify our
circuit for calculating
$\pp(\calf|D)$
is to give a circuit for
generating $\ket{s}$.

See
Fig.\ref{fig-halfmoon}
where a halfmoon vertex that we will use in the next figure is defined. Controlled gates
with one of these halfmoon
vertices at the bottom should
be interpreted as two gates, one with
a $P_0$ control replacing the halfmoon,
and one with a $P_1$ control replacing it. The one with the $P_1$ control
should have at the top a $R_y$ rotation
and the one with the $P_0$ control
should have at the top a Hadamard matrix. All controls acting on qubits other than the qubits at the very top and very bottom are kept the same.

A circuit for generating
$\ket{s}$ is given by
Fig.\ref{fig-qJen-ckt}. The $V_0^{(\lam)}$ matrices
 used in Fig.\ref{fig-qJen-ckt}.
    are defined in Ref.\cite{qSym}.
Fig.\ref{fig-qJen-ckt}
is equivalent to saying that

\beq
\ket{s}_{\mu,\nu,\omega}=
\sigma_X(\omega)^{
\pi(\beta)
\pi(\alpha)}
\frac{1}{\sqrt{2}}
\left[
\begin{array}{l}
T_1(\alpha)\ket{0^3}_{\alpha}
\\
\ket{0^{12}}_\beta
\\
\ket{1}_\gamma
\\
\ket{1}_{\mu_0}
\\
\ket{1}_\omega
\end{array}
+
\begin{array}{l}
T_0(\alpha)\ket{0^3}_{\alpha}
\\
\ket{0^{12}}_\beta
\\
\ket{0}_\gamma
\\
\ket{0}_{\mu_0}
\\
\ket{1}_\omega
\end{array}
\right]
\;.
\eeq

\begin{claim}\label{cl-qJen-ckt}

\beq
\ket{s}_{\mu,\nu,\omega}=
\begin{array}{c}
z_1 \ket{\psi_1}_\mu
\\
\ket{1}_\nu
\\
\ket{0}_\omega
\end{array}
+
\begin{array}{c}
z_0 \ket{\psi_0}_\mu
\\
\ket{0}_\nu
\\
\ket{0}_\omega
\end{array}
+
\begin{array}{c}
\ket{\chi}_{\mu,\nu}
\\
\ket{1}_\omega
\end{array}
\;,
\eeq
for some unnormalized state
$\ket{\chi}_{\mu,\nu}$,
where
\beq
\begin{array}{|c|c|}
\hline
\ket{\psi_1}_\mu=
\begin{array}{l}
\ket{1^3}_\alpha
\\
\ket{1}_{\mu_0}
\end{array}
&
\ket{\psi_0}_\mu=
\begin{array}{l}
\ket{1^3}_\alpha
\\
\ket{0}_{\mu_0}
\end{array}
\\
\ket{1}_\nu=
\left[
\begin{array}{r}
\ket{0^3}_{\beta_{;0}}
\\
\ket{0^6}_{\beta_{;1}}
\\
\ket{0^3}_{\beta_{;2}}
\\
\ket{1}_{\gamma}
\end{array}
\right]
&
\ket{0}_\nu=
\left[
\begin{array}{r}
\ket{0^3}_{\beta_{;0}}
\\
\ket{0^6}_{\beta_{;1}}
\\
\ket{0^3}_{\beta_{;2}}
\\
\ket{0}_{\gamma}
\end{array}
\right]
\\
\hline
\end{array}
\;,
\eeq

\begin{subequations}\label{eq-z0-z1-claim}
\beq
z_1 = \frac{\epsilon}{\sqrt{2}}
\sum_\sigma
\overbrace{
\left\{\sin(
\theta_{2^\sigma|\{1^\sigma, 0^\sigma\}})
\sin(\theta_{1^\sigma|0^\sigma})
\sin(\theta_{0^\sigma})\right\}}^
{=h_{2^\sigma|\{1^\sigma, 0^\sigma\}}
h_{1^\sigma|0^\sigma}
h_{0^\sigma}}
\;,
\eeq

\beq
z_0 = \frac{\epsilon n!}{\sqrt{2^{n+1}}}
\;,
\eeq
\end{subequations}

\beq
\frac{|z_1|}{|z_0|} = \sqrt{\frac{P(1)}{P(0)}}
\;.
\label{eq-z1-z0}
\eeq
\end{claim}
\proof

Recall that for any
quantum systems $\alpha$ and $\beta$,
any
unitary operator $U(\beta)$
and any
projection operator $\pi(\alpha)$,
one has

\beq
U(\beta)^{\pi(\alpha)}=
(1-\pi(\alpha)) + U(\beta)\pi(\alpha)
\;.
\label{eq-u-pi-id}
\eeq
Applying identity Eq.(\ref{eq-u-pi-id}) with $U=\sigma_X(\omega)$
yields:

\beqa
\ket{s} &=&
\sigma_X(\omega)^{\pi(\beta)\pi(\alpha)}\ket{s'}
\\
&=&
\sigma_X(\omega)\pi(\beta)\pi(\alpha)\ket{s'}
+
\begin{array}{l}
\ket{\chi}_{\mu,\nu}
\\
\ket{1}_\omega
\end{array}
\;.
\eeqa

Note that
$T_1(\alpha)$ is a sum of $n^n$ terms,
which is more than $n!$ terms,
but the controls in those
$n^n$ terms
together with the act
of taking the matrix element
between $\bra{1^3}$ and $\ket{0^3}$,
reduces the number of
summed over terms to $n!$:

\beqa
\bra{1^3}T_1(\alpha)\ket{0^3}
&=&
\epsilon \sum_{\sigma \in Sym_n}
\bra{1^3}
\begin{array}{r}
\exp\left[
-i\sigma_Y(\alpha_{2^\sigma})
\theta_{2^\sigma|\{1^\sigma, 0^\sigma\}}
\right]
\\
\exp\left[
-i\sigma_Y(\alpha_{1^\sigma})
\theta_{1^\sigma|0^\sigma}
\right]
\\
\exp\left[
-i\sigma_Y(\alpha_{0^\sigma})
\theta_{0^\sigma}
\right]
\end{array}
\ket{0^3}
\\
&=&
\epsilon \sum_{\sigma}
\begin{array}{r}
\sin(
\theta_{2^\sigma|\set{1^\sigma, 0^\sigma}}
)
\\
\sin(
\theta_{1^\sigma|0^\sigma}
)
\\
\sin(
\theta_{0^\sigma}
)
\end{array}
\;.
\eeqa
Likewise,
\beqa
\bra{1^3}T_0(\alpha)\ket{0^3}
&=&
\epsilon \sum_{\sigma \in Sym_n}
\bra{1^3}H(\alpha_2)H(\alpha_1)H(\alpha_0)
\ket{0^3}
\\
&=&
\frac{\epsilon n!}
{\sqrt{2^n}}
\;.
\eeqa
\qed

To draw
the circuit of Fig.\ref{fig-qJen-ckt},
especially for $n$ much larger than 3,
requires that one know how to enumerate
all possible combinations of $\ell$
elements from a set of $n$ elements.
A simple algorithm for doing this is known (Ref.\cite{flamig}). It's based on a careful study of
the pattern in simple examples such as this one:

Choose 3 elements out of the set $\{0,1,2,3,4\}$:
\beq
\begin{array}{|c|c|c|c|c|}
\hline
0&1&2& &
\\  \hline
0&1& &3&
\\  \hline
0&1& & &4
\\  \hline
0& &2&3&
\\  \hline
0& &2& &4
\\  \hline
0& & &3&4
\\  \hline
 &1&2&3&
\\  \hline
 &1&2& &4
\\  \hline
 &1& &3&4
\\  \hline
 & &2&3&4
 \\ \hline
\end{array}
\;
\eeq

\begin{figure}[h]
    \begin{center}
    \epsfig{file=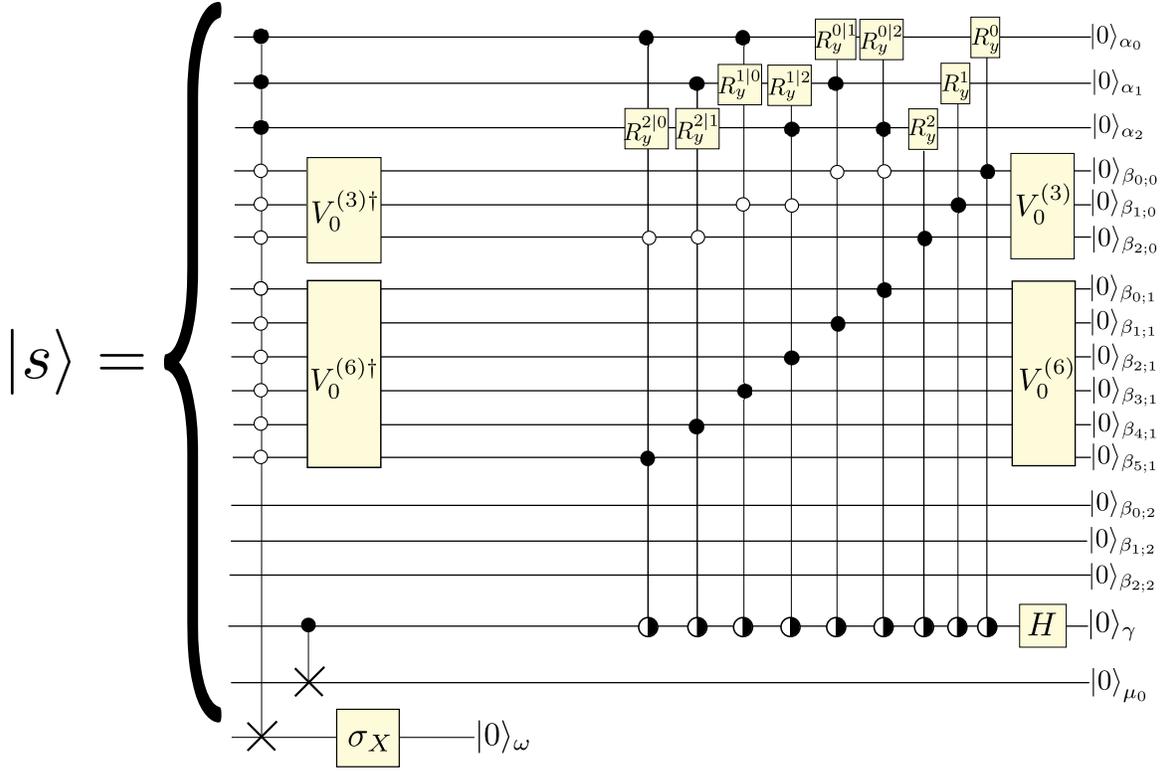, width=6in}
    \caption{Circuit for
    generating $\ket{s}$
    used in AFGA to calculate
    $\pp(\calf| D)$
    for ordered modular model.
    This is a simplification of
    Fig.\ref{fig-qJen-ckt},
     assuming that the in-degree of all nodes
    is smaller or equal to one.
    }
    \label{fig-qJen-ckt-restrict}
    \end{center}
\end{figure}

A more serious problem with using the
circuit of Fig.\ref{fig-qJen-ckt} for large $n$ is that
as Eq.(\ref{eq-exp-n2b}) indicates,
the number of $\beta$ qubits grows
exponentially with $n$ so the circuit
Fig.\ref{fig-qJen-ckt}
is too expensive for large $n$'s.
However, one can make an
assumption which doesn't seem too
restrictive, namely that the
in-degree (number of parent nodes) $\ell$ of all nodes of
the graph $G$ is $\leq \ell_{max}$,
where
the bound $\ell_{max}$  does not grow with $n$.
Define

\beq
N_2'(\beta)=
 \sum_{\ell=0}^{\ell_{max}}
N_2(\beta_{;\ell})
=\calo(n)
\;
\label{eq-o-est}
\eeq
and

\beq
\epsilon'=
\frac{1}{
\prod_{\ell=0}^{\ell_{max}}
N_2(\beta_{;\ell})
}
\;.
\eeq
The order estimate for
$N'_2(\beta)$
given by Eq.(\ref{eq-o-est}) can
be proven using Stirling's approximation.

For example, consider
Fig.\ref{fig-qJen-ckt}.
If $\ell\leq \ell_{max}=1$ for that figure, then we can omit
all the $\beta_{;2}$ qubits, and the $R_y^{a|\set{b,c}}$ rotations
for $a,b,c\in\set{0,1,2}$.
In other words,
Fig.\ref{fig-qJen-ckt}
can be simplified to
Fig.\ref{fig-qJen-ckt-restrict}.
Claim \ref{cl-qJen-ckt} still holds
if we replace
$h(2^\sigma|\{1^\sigma,0^\sigma\})$ by
1
and
$\epsilon$ by $\epsilon'$
in Eqs.(\ref{eq-z0-z1-claim}).

\end{document}